\documentclass[preprintnumbers,
nofootinbib,
notitlepage,
amsmath,amssymb,
]{revtex4-1}

\usepackage[dvipdfmx]{graphicx}

\usepackage{graphicx}
\usepackage{dcolumn}
\usepackage{bm}

\makeatletter
\def\slash#1{{\mathpalette\c@ncel{#1}}} 
\makeatother

\newcommand\beq{\begin{eqnarray}}
\newcommand\eeq{\end{eqnarray}}

\newcommand{\Slash}[1]{{\ooalign{\hfil/\hfil\crcr$#1$}}}

\newcommand{\nn}{\nonumber \\}

\begin{document}

\preprint{YITP-16-125}

\title{Single spin asymmetry in forward $pA$ collisions II: Fragmentation contribution }

\author{Yoshitaka Hatta$^{\rm a}$}
\author{Bo-Wen Xiao$^{\rm b}$}
\author{Shinsuke Yoshida$^{\rm c}$}%
\author{Feng Yuan$^{\rm d}$}

\affiliation{\vspace{3mm}
  $^{\rm a}$Yukawa Institute for Theoretical Physics, Kyoto University, Kyoto 606-8502, Japan \\
  $^{\rm b}$ Key Laboratory of Quark and Lepton Physics (MOE) and Institute of Particle Physics, Central China Normal University, Wuhan 430079, China \\
 $^{\rm c}$ Theoretical Division, Los Alamos National Laboratory, Los Alamos, NM 87545, USA \\ 
  $^{\rm d}$ Nuclear Science Division, Lawrence Berkeley National Laboratory, Berkeley, CA 94720, USA \\
}%

\date{\today}

\begin{abstract}
\vspace{4mm}
We compute the twist-three fragmentation contribution to the transverse single spin asymmetry 
(SSA) in light hadron production $p^\uparrow p\to hX$ and $p^\uparrow A\to hX$  including the 
gluon saturation effect in the unpolarized nucleon/nucleus. Together with the results in our previous paper,
this completes the full evaluation of the SSA in this process in the ``hybrid" formalism. 
We argue that the dependence of SSAs on the atomic mass number in the forward 
region can elucidate the relative importance of the soft gluon pole contribution from the twist-three 
quark-gluon-quark correlation in the polarized nucleon and the twist-three fragmentation contribution 
from the final state hadron.

\end{abstract}

\pacs{Valid PACS appear here}
\maketitle




\section{Introduction}

Single-transverse spin asymmetries (SSAs) in inclusive hadron 
production in nucleon-nucleon scattering, $p^{\uparrow} p\to hX$, remain
one of the long standing puzzles in hadron physics. 
In recent years, the physicists at the Relativistic Heavy Ion Collider (RHIC)
have planned and explored the SSAs in the forward hadron production in nucleon-nucleus
collisions, $p^{\uparrow} A\to hX$~\cite{Aschenauer:2015eha,Heppelmann:2016siw}. This will not only provide additional
information on the underlying mechanism for the SSA phenomena,
but also help us understand the small-$x$ saturation of the 
gluon distributions in large nuclei. 

In a previous paper \cite{Hatta:2016wjz}, we have computed the SSA 
of light hadrons in proton-nucleus collisions $p^\uparrow A \to hX$ 
including the small-$x$ gluon saturation effect in the nucleus. We adopted the 
so-called hybrid approach \cite{Schafer:2014zea,Zhou:2015ima} where the 
collinear twist-three Efremov-Teryaev-Qiu-Sterman (ETQS) functions \cite{Efremov:1984ip,Qiu:1991pp} 
are used on the polarized proton side and the unintegrated ($k_T$-dependent) gluon distribution 
is used on the nucleus side. We find that leading terms
in the forward region come from the soft-gluon pole contributions
of the twist-three ETQS matrix elements in the transversely polarized 
nucleon. In particular, the so-called derivative term will dominate the SSA in the forward region. From this we concluded that the asymmetry $A_N$ does not depend on the saturation scale of the nucleus. Of course, for a complete
evaluation in this hybrid approach, we also have to take into
account the twist-three fragmentation function contributions. (See, also, \cite{Kovchegov:2012ga}.) The goal of this paper is to carry out this part of the calculation.

In the purely collinear framework, 
the twist-three fragmentation function contribution has been first studied in  \cite{Kang:2010zzb} 
and completed in \cite{Metz:2012ct} (see a recent review \cite{Pitonyak:2016hqh}). 
The gauge and Lorentz invariance of the result has been recently established   \cite{Kanazawa:2015ajw}.  
In the forward region of $p^\uparrow A$ collisions, the saturation effect in the nucleus becomes important. 
The effect of saturation on the fragmentation contribution has been so far considered only in the 
$k_T$-factorization approach \cite{Kang:2011ni} which involves the Collins function \cite{Collins:1992kk}.
However, in the Sivers-type contribution, we have found \cite{Hatta:2016wjz} that the 
$k_T$-factorization approach \cite{Boer:2006rj}  misses the dominant derivative term. 
Whether this happens also in the fragmentation contribution is phenomenologically important, 
especially in view of the recent claim \cite{Kanazawa:2014dca} that the SSA in  
$p^\uparrow p \to hX$ is completely dominated by the `genuine twist-three' 
fragmentation function, with both the Sivers and Collins contributions playing 
only a minor role. However, the assumption of a large genuine twist-three fragmentation 
function made in \cite{Kanazawa:2014dca} has not been tested yet  
because there are no other available experimental data sensitive to this function. 
In this paper, we show  that the dependence of SSA on the mass number of the 
nucleus, as recently measured at RHIC \cite{Heppelmann:2016siw}, can be  such a test.

In the hybrid formalism,\footnote{The twist-three contribution from the unpolarized nucleon/nucleus
in the current kinematics is suppressed in the small-$x$ calculations, and neglected in this paper.} 
the single transverse spin-dependent cross section  can be schematically written as 
\begin{eqnarray}
E_h\frac{d^3\Delta\sigma (p^\uparrow A\to hX)}{d^3\vec{P}_{h}}&=&\epsilon^{ij}S_{T i}P_{hj}\int_{x_F}\frac{dz}{z^2}
\Bigl\{
D_{h/q}(z) G_F(x_p,x_p)\otimes F(x_g,P_{hT}/z) \nonumber\\
&&  \qquad \quad +h_{1}(x_p)\hat{H}(z)\otimes F(x_g,P_{hT}/z)\Bigr\}\,. \label{first}
\end{eqnarray}
The first term is what we have calculated in Ref.~\cite{Hatta:2016wjz}, and the second term is the 
object of this paper. In the above equation, $S_T$ represents the traverse polarization vector of the projectile, 
 $P_{hT}$ is the transverse momentum of the final state hadron.
$h_{1}(x_p)$ is the collinear leading-twist quark transversity distribution function and $D(z)$ is 
the leading-twist fragmentation function,
whereas $G_F(x_p,x_p)$ and $\hat H(z)$ represent the twist-three ETQS distribution from the polarized
nucleon and the twist-three fragmentation function, respectively. 
The small-$x$ saturation physics is encoded in the unintegrated gluon distribution (or the dipole gluon distribution)  $F(x_g,k_T)$.
Although both of the contributions in (\ref{first}) are classified as twist-three in the collinear
approach, the underlying mechanisms are different.
The twist-three terms associated with the incoming polarized nucleon
comes from the initial/final state interaction effects which are necessary to 
generate a phase from the pole contributions. On the other hand, the 
twist-three fragmentation function contributions do not need a phase 
from the scattering amplitudes as we will show in the following calculations. 
Because of this difference, we expect that the two contributions depend differently  
on the saturation scale (or the atomic mass number).

The rest of the paper is organized as the following.
In Section II, we compute the twist-three fragmentation contribution in the hybrid approach without including the saturation effect in the target.  We explicitly check that, at large-$P_{hT}$, our result agrees with the previous result obtained in the collinear factorization framework \cite{Metz:2012ct}. 
We then include the saturation effects
and present the complete formula in Section III. Finally in Section IV, we 
discuss the phenomenological consequences of our result.

\section{Fragmentation contribution to SSA}

\begin{figure}[t]
\begin{center}
  \includegraphics[width=15cm,height=4.6cm]{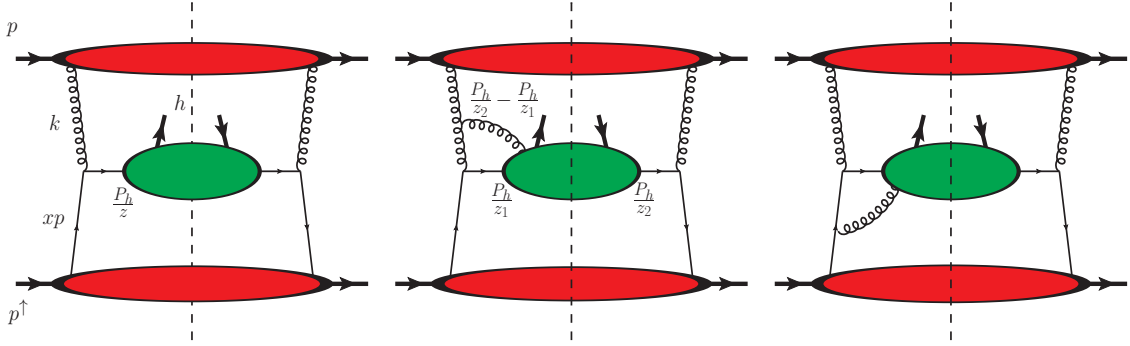}
\end{center}
 \caption{ Fragmentation contribution to single spin asymmetry in the hybrid approach. The left diagram represents the first two terms in (\ref{yu}). The middle and the right diagrams represent the last term in (\ref{yu}).    } \label{fig1}
\end{figure}

In this section we compute the fragmentation contribution to SSA in the hybrid approach in the `dilute' limit, i.e., without including the saturation effect in the target.
  Our starting point is Eq.~(54) of Ref.~\cite{Kanazawa:2013uia} which was derived for semi-inclusive DIS (SIDIS) $e p^\uparrow \to ehX$ but is valid also for $p^\uparrow p \to hX$. 
The  spin-dependent part of the cross section is
\beq
E_h\frac{d\sigma^{frag}}{d^3\vec{P}_{h}}&=&\frac{1}{4s (2\pi)^3} \Biggl\{ \int \frac{dz}{z^2} {\rm Tr}[\Delta(z) S(z)] + \int \frac{dz}{z^2}
{\rm Im} {\rm Tr} \left[\Delta^\alpha_\partial (z) \frac{\partial S(K)}{\partial K^\alpha}\right]_{K=\frac{P_h}{z}} \nn 
&& - \int \frac{dz_1dz_2}{z_1^2z_2^2} P\left(\frac{1}{1/z_2-1/z_1}\right) {\rm Im} {\rm Tr} [\Delta_F^\alpha (z_1,z_2)(S_\alpha^L(z_1,z_2)+S_\alpha^R(z_1,z_2))] \Biggr\}\,,
\label{yu}
\eeq
where $P_h^\mu$ is the momentum of the measured hadron species $h$ whose mass is neglected $P_h^2=2P_h^+P_h^- -P_{hT}^2=M_h^2\approx 0$. The momenta of the polarized and unpolarized protons are denoted by $p^\mu$ and $q^\mu$, respectively. The center-of-mass energy is then $s\approx 2p^+q^-$.  $\Delta$'s describe the fragmentation process into $h$, and $S$'s represent the rest of the cross section. 
We shall be interested in the forward region $P_h^+ \gg P_{hT} \gg P_h^-$ and keep only the leading contributions in $P_{hT}/P_h^+$. In this kinematics, $S$ and $S^{L}$ are depicted in the first and the last two diagrams of Fig.~\ref{fig1}, respectively. ($S^R$ is the mirror image of $S^L$.) In our approach, the transverse momentum of the final state hadron $P_{hT}$ comes from the intrinsic transverse momentum of the small-$x$ gluon from the unpolarized target. This is why we only consider $2\to 1$ scattering instead of $2 \to 2$ scattering. 

The twist-three fragmentation functions are contained in $\Delta$'s as
\beq
\Delta (z) = \frac{M}{2z} \sigma_{\lambda \alpha} i\gamma_5 \epsilon^{\lambda\alpha w P_h}\hat{e}_{\bar{1}}(z) + \cdots\,,\label{2}
\eeq
\beq
\Delta_{\partial}^\alpha = \frac{M}{2} \gamma_5 \frac{\Slash P_h}{z} \gamma_\lambda  \epsilon^{\lambda \alpha w P_h} \tilde{e}(z) + \cdots\,,  \label{3}
\eeq
\beq
\Delta_{F}^\alpha(z_1,z_2) = \frac{M}{2} \gamma_5 \frac{\Slash P_h}{z_2} \gamma_\lambda\epsilon^{\lambda \alpha w P_h}\hat{E}_F(z_1,z_2) + \cdots\,, \label{4}
\eeq
where $M$ is the proton mass.   We use the conventions $D^\mu=\partial^\mu -igA^\mu$,   $\gamma_5=i\gamma^0\gamma^1\gamma^2\gamma^3$ and $\epsilon^{\lambda \alpha w P_h}\equiv \epsilon^{\lambda\alpha\rho\sigma}w_\rho P_{h\sigma}$ with  $\epsilon_{0123}=+1$. The two-dimensional antisymmetric tensor $\epsilon^{ij}$  is defined as $\epsilon^{12}=-\epsilon^{21}=1$ so that $\epsilon^{+-ij}=\epsilon^{ij}$. (We use Latin letters $i,j,l=1,2$ for  transverse indices.) 
$w^\mu$ is a vector which satisfies the conditions $P_h\cdot w=1$ and $w^2=0$. Explicitly, 
\beq
(w^+,w^-,w^i) = \frac{1}{2E_h^2}(P_h^-,P_h^+,-P_h^i) \approx   \frac{1}{(P_h^+)^2}(P_h^-,P_h^+,-P_h^i)
\eeq
The largest component is $w^- \approx 1/P_h^+$. 
 The three functions in (\ref{2})-(\ref{4}) are not totally independent. They satisfy the relation
\beq
\frac{\hat{e}_{\bar{1}}(z)}{z} -{\rm Im} \tilde{e}(z) = \int \frac{dz'}{z'^2} P\frac{1}{1/z'-1/z} {\rm Im} \hat{E}_F(z',z)\,. \label{ident}
\eeq  
The relevant distribution function for the transversely polarized proton is the transversity distribution $h_1(x)$ 
\beq
\langle p|\psi \bar{\psi} |p\rangle = \frac{1}{8}\langle \bar{\psi}i\gamma_5\sigma^{\mu\nu}\psi\rangle  i\gamma_5\sigma_{\mu\nu} + \cdots = -\frac{p^+S_{Ti}}{2} \int dx h_1(x) i\gamma_5\sigma^{-i}+\cdots\,,  \label{lead}
\eeq
 where $\vec{S}_T$ is the transverse spin vector normalized as $\vec{S}_T^2 =1$.

\subsection{First term}

Let us calculate the three terms in (\ref{yu}) one by one. The integrand of the first term reads 
\beq
 {\rm Tr}[\Delta(z) S(z)]  &=& -g^2C_F \frac{Mp^+S_{T i}}{4z} \epsilon^{\lambda\alpha w P_h} \hat{e}_{\bar{1}}(z) \int dx h_1(x)   \int d^3k {\rm Tr} [ i\gamma_5\sigma^{-i} \gamma^\nu \sigma_{\lambda\alpha}i\gamma_5 \gamma^\mu ] \nn 
&& \times \frac{ \langle q| A_\mu(k)A_\nu(-k)|q\rangle}{N_c^2-1} (2\pi)^4\delta^{(4)} \left(xp+k-\frac{P_h}{z}\right) \nn 
&=& -(2\pi)^4 g^2 \frac{MS_{T i}}{8N_c} \epsilon^{\lambda\alpha w P_h} \frac{\hat{e}_{\bar{1}}(z)}{z}  h_1(x)   {\rm Tr} [ i\gamma_5\sigma^{-i} \gamma^\nu \sigma_{\lambda\alpha}i\gamma_5 \gamma^\mu ]   \langle q| A_\mu(k)A_\nu(-k)|q\rangle \,,  \label{fi}
\eeq
where $C_F=\frac{N_c^2-1}{2N_c}$ and $k^\mu=(0,k^-,\vec{k}_T)$. The momentum conserving  delta function fixes the components of $P_h^\mu$ as
\beq
P_h^+ = xzp^+ \equiv x_F p^+ \,, \qquad P_h^- = zk^- \,, \qquad \vec{P_h}_T = z\vec{k}_T\,.
\eeq
Spin-dependent cross sections are often measured at fixed $x_F$.  In the forward region in which we are interested, $x_F \approx 1$.  
Working out the trace of gamma matrices, we get 
\beq
&&\epsilon^{\lambda\alpha w P_h}{\rm Tr} [ i\gamma_5\sigma^{-i} \gamma^\nu \sigma_{\lambda\alpha}i\gamma_5 \gamma^\mu ]A_\mu(k) A_\nu(-k) 
 \nn  
&&= -8 \Bigl(-\epsilon^{\alpha -wP_h} ( A^i A_\alpha + A_\alpha A^i ) + \epsilon^{\lambda i wP_h} (A^- A_\lambda + A_\lambda A^- ) -\epsilon^{-i wP_h} A^\mu A_\mu \Bigr) \nn 
&& \approx \frac{-8}{P_h^+} \left( \epsilon^{lj}P_{hj} (A^i A_l + A_l A^i) + \epsilon^{li } \Bigl(P_{hl} (2A^- A^- + A^j A_j) +P_h^+ (A^- A_l + A_l A^-)  \Bigr) \right) \,. \label{9}
\eeq
One might be puzzled by this complicated expression which cannot be rewritten as a gauge invariant combination of $F^{\mu\nu}(k)=i(k^\mu A^\nu- k^\nu A^\mu)+{\mathcal O}(g)$. In fact, the other terms in (\ref{yu}) also give similar, non-gauge-invariant terms, and the identity (\ref{ident}) is needed to check whether the sum is gauge invariant \cite{Kanazawa:2015ajw}. However, this is beyond the scope of this work. A simple counting argument $A^- \sim q^- \sim P_h^+ \gg A^i\sim P_{hT}$ shows that the whole expression  (\ref{9}) is  subleading by a factor $(P_{hT}/P_h^+)^2$ compared to what we shall keep in the end,  and at this subleading level diagrams other than those in Fig.~\ref{fig1} will come into play. We thus  simply  ignore (\ref{9})  for the present purpose.

\subsection{Second term} 

The second term in (\ref{yu}) is evaluated as 
\beq
 {\rm Im} {\rm Tr} \left[\Delta^\alpha_\partial (z) \frac{\partial S(K)}{\partial K^\alpha}\right]_{K=\frac{P_h}{z}}  
 &=& -g^2 (2\pi)^4 \frac{MS_{Ti} }{2N_c} \epsilon^{\lambda \alpha w P_h} {\rm Im}\,  \tilde{e}(z)  \nn 
&\times& \frac{\partial}{\partial K^\alpha}  \left[ h_1(K^+/p^+)\frac{1}{4}  {\rm Tr} [i\gamma_5 \sigma^{-i} \gamma^\nu \gamma_5 \frac{\Slash P_h}{z} \gamma_\lambda \gamma^\mu  ]  \langle A_\mu(\tilde{K})A_\nu(-\tilde{K})\rangle  \right]_{K=\frac{P_h}{z}}\,, \label{ut}
\eeq
 where we introduced the notation $\tilde{K}^\mu=(0,K^-,\vec{K}_T)$. 
We use the trick 
\beq
&&\frac{\partial}{\partial K^\alpha}  \left[ h_1(K^+/p^+)  {\rm Tr} [i\gamma_5 \sigma^{-i} \gamma^\nu \gamma_5 \frac{\Slash P_h}{z} \gamma_\lambda \gamma^\mu  ]   A_\mu(\tilde{K})A_\nu(-\tilde{K})\right]_{K=\frac{P_h}{z}} \label{can}  \nn
&& = \frac{\partial}{\partial K^\alpha}  \Bigl[ h_1(K^+/p^+)  {\rm Tr} [i\gamma_5 \sigma^{-i} \gamma^\nu \gamma_5 \Slash K \gamma_\lambda \gamma^\mu  ]  A_\mu(\tilde{K})A_\nu(-\tilde{K}) \Bigr]_{K=\frac{P_h}{z}}  \nn 
&& \qquad -  h_1(x)  {\rm Tr} [i\gamma_5 \sigma^{-i} \gamma^\nu \gamma_5 \gamma_\alpha \gamma_\lambda \gamma^\mu  ]   A_\mu(k)A_\nu(-k)  \label{tri} \,. 
\eeq
The second term on the right hand side has exactly the same $\gamma$-matrix structure as in (\ref{fi}).\footnote{One can replace $\gamma_\alpha \gamma_\lambda \to \frac{1}{2}[\gamma_\alpha,\gamma_\lambda]$ due to the presence of $\epsilon^{\lambda\alpha wP_h}$.} It is thus subleading in energy and can be dropped.\footnote{Incidentally, if we add this term to (\ref{fi}), we get the combination  $\frac{\hat{e}_{\bar{1}}}{z} -{\rm Im} \, \tilde{e}(z)$ 
 which appears in the identity (\ref{ident}). }
As for the first term in (\ref{tri}), we find
\beq
&& \frac{1}{4} {\rm Tr} [i\gamma_5 \sigma^{-i} \gamma^\nu \gamma_5 \Slash K \gamma_\lambda  \gamma^\mu  ] A_\mu A_\nu 
 \nn 
 && = \delta^i_\lambda (K\cdot A A^- + A^- K\cdot A -K^- A^\mu A_\mu) -\delta^-_\lambda( K\cdot A A^i + A^i K\cdot A -K^i A^\mu A_\mu)  \nn && \qquad +  K_\lambda( A^- A^i -A^i A^-) + A_\lambda(K^-A^i -K^i A^-) + (K^-A^i -K^i A^-)A_\lambda\,.
\eeq
The dominant term is $\sim \delta^i_\lambda K^+A^-A^-$ which combines with other terms to form the   gauge invariant operator\footnote{Note that terms proportional to $K^2$ and $K_\lambda$ can be omitted.  If the $K^\alpha$-derivative in (\ref{can}) acts on $K_\lambda$, it gives $g_{\alpha\lambda}$ and vanishes when contracted with $ \epsilon^{\lambda \alpha w P_h}$. If the derivative does not act on $K_\lambda$, then after setting $K_\lambda = P_{h\lambda}/z$ we get zero $P_{h\lambda}\epsilon^{\lambda \alpha wP_h}=0$. Similarly, if the derivative acts on $K^2$, it gives $K_\alpha$ and  vanishes after replacing $K_\alpha \to P_{h\alpha}/z$. If the derivative does not act on $K^2$, then again it vanishes because $K^2 \to P_h^2/z^2 =0$.} 
\beq
F^{-\mu}F^-_{\ \mu} &=& K^- \Bigl( K^- A^\mu A_\mu  - (\tilde{K}\cdot A A^- + A^- \tilde{K}\cdot A) \Bigr)+ \tilde{K}^2  A^-A^-  +{\mathcal O}(g) \nn 
&=& K^- \Bigl( K^- A^\mu A_\mu  - (K\cdot A A^- + A^- K\cdot A) \Bigr)+ K^2  A^-A^-  +{\mathcal O}(g) 
\,.
\eeq
To twist-two accuracy, we only keep this term and  
 use  
\beq
\frac{\langle F^{-i}F^{-j}\rangle}{K^-} = \frac{K^i K^j}{K_T^2}  G(x_g,K_T)\,, \qquad 
\frac{\langle F^{-\mu}F^{-}_{\ \mu}\rangle}{K^-} = - G(x_g,K_T)\,, \qquad (x_g=K^-/q^-) \label{two}
\eeq
where $G$  is the unintegrated gluon distribution of the unpolarized proton. 
The $K$-derivative can be decomposed  as
\beq
\epsilon^{i+wP_h}\frac{\partial}{\partial K^+} +\epsilon^{i-wP_h}\frac{\partial}{\partial K^-} +  \epsilon^{ijwP_h}\frac{\partial}{\partial K^j}
\approx \epsilon^{ij} \left[ \frac{P_{hj}}{P_h^+} \left( \frac{\partial}{ \partial K^+}- \frac{\partial}{ \partial K^-}\right) + \frac{\partial}{\partial K^j} \right]\,.
\eeq
The $K^+$-derivative can be safely neglected. However, the $K^-$-derivative should be kept since $1/K^-\sim z/P^-_h$ is large. This can be combined with the $K_T$-derivative as
\beq
\left.\left(-\frac{P_{hj}}{P_h^+} \frac{\partial}{\partial K^-} + \frac{\partial}{\partial K^j}\right) G\left(x_g=\frac{K^-}{q^-},K_T \right) \right|_{K=\frac{P_h}{z}} 
= \frac{d}{d(P_h^j/z)} G\left(x_g=\frac{P_{hT}^2}{xz^2s}, \frac{P_{hT}}{z}\right)\,.
\eeq
We thus arrive at
\beq
 {\rm Im} {\rm Tr} \left[\Delta^\alpha_\partial (z) \frac{\partial S(K)}{\partial K^\alpha}\right]_{K=\frac{P_h}{z}}   =-  \frac{g^2M}{2N_c} (2\pi)^4   h_1(x) {\rm Im}\,  \tilde{e}(z)  S_{T i}  \epsilon^{ij}   \frac{d}{d(P_h^j/z)} G\left(x_g=\frac{P_{hT}^2}{xz^2s}, \frac{P_{hT}}{z}\right)    \,. \label{cro}
\eeq

\subsection{Third term} 

The last term in (\ref{yu}) is the `genuine twist-three' contribution
\beq
&&  {\rm Im} {\rm Tr} [\Delta_F^\alpha(z_1,z_2)(S_\alpha^L(z_1,z_2)+S_\alpha^R(z_1,z_2))] =g^2M\frac{(2\pi)^4}{4z_2}  h_1(x) S_{T i}  \epsilon^{ \lambda \alpha wP_h} {\rm Im}\hat{E}_F (z_1,z_2)  \nn && \times \Biggl\{ 
\frac{N_c}{2}\frac{ {\rm Tr}\left [ i\gamma_5 \sigma^{-i} \gamma^\nu \gamma_5 \Slash P_h \gamma_\lambda \gamma^\beta \right] }{(k+P_h(1/z_1-1/z_2))^2}  \Bigl(\delta^\mu_{\alpha}(\frac{P_h}{z_1}-\frac{P_h}{z_2} -k)_\beta - g_{\alpha \beta}(k+2\frac{P_h}{z_1}-2\frac{P_h}{z_2})^\mu + \delta^\mu_{\beta}(2k+\frac{P_h}{z_1}-\frac{P_h}{z_2})_\alpha \Bigr) \nn
&& \qquad   -\frac{1}{2N_c }  {\rm Tr}\left [ i\gamma_5 \sigma^{-i} \gamma^\nu \gamma_5\Slash P_h \gamma_\lambda \gamma^\mu  \frac{ x\Slash p + \Slash P_h (1/z_1 -1/z_2)  }{(xp+ P_h(1/z_1-1/z_2))^2}\gamma_\alpha \right] + (\mu \leftrightarrow \nu)  \Biggr\} \frac{\langle A_\mu (k) A_\nu(-k)\rangle}{N_c^2-1}  \,.  \label{last}
\eeq
The two terms correspond to the middle and right diagrams of Fig.~\ref{fig2} and have different dependence on $N_c$.
Let us first look at the ${\mathcal O}(1/N_c)$ contribution.  The quark propagator contains two terms, $x\Slash p$ and $\Slash P_h(1/z_1-1/z_2)$. The former gives 
\beq
&& xp^+ \epsilon^{ \lambda \alpha wP_h} {\rm Tr} \left [ i\gamma_5\sigma^{-i}  \gamma^\nu \gamma_5 \Slash P_h \gamma_\lambda \gamma^\mu  \gamma^- \gamma_\alpha + (\mu \leftrightarrow \nu) \right] \langle A_\mu A_\nu \rangle  \label{34}  \\ 
&& = \frac{16P_h^+}{z_2}\epsilon^{ \lambda - wP_h} \left\langle  A_\lambda (P_h^-A^i-P_h^i A^-) + (P_h^-A^i-P_h^i A^-) A_\lambda+\delta^i_\lambda(A^- P_h\cdot A +P_h\cdot A A^-  -P_h^- A^\mu A_\mu) \right\rangle\,. \nonumber  
\eeq
Using relations such as
\beq
A^- P_h\cdot A +P_h\cdot A A^-  -P_h^- A^\mu A_\mu
&=&\frac{z_2 }{k^-}  \left( k_{T}^2 A^-A^- + k^-k_{i} (A^-A^i +A^iA^-) -(k^-)^2A^iA_i\right) 
\nn
&=& -\frac{z_2}{k^-} F^{-\mu}F^-_{\ \mu} \,, 
\eeq
and (\ref{two}), we can rewrite  (\ref{34}) in the form 
\beq
16\frac{P_h^+}{k^-} \epsilon^{ \lambda - wP_h} \left\langle F^-_{\ \lambda}F^{-i} + F^{-i}F^-_{\ \lambda} -\delta^i_\lambda F^{-\mu}F^-_{\ \mu} \right\rangle &\approx& 16P_{h}^+\epsilon^{j-+l}w^-P_{hl} \left(\frac{2k_jk^i}{k_T^2} + \delta^i_j\right)G(x_g,k_T)  \nn &=& -16 \epsilon^{i j}P_{hj} G(x_g,k_T)\,.
\eeq
The other term $\Slash P_h(1/z_1-1/z_2)$ can be evaluated as  
\beq
  \epsilon^{ \lambda \alpha wP_h} {\rm Tr} \left [ i\gamma_5\sigma^{-i}  \gamma^\nu \gamma_5 \Slash P_h \gamma_\lambda \gamma^\mu  \Slash P_h \gamma_\alpha \right]
 &=&-P_h^\mu \epsilon^{ \lambda \alpha wP_h} {\rm Tr} \left [ \gamma^-\gamma^i \gamma^\nu (\gamma_\lambda\Slash P_h   \gamma_\alpha - \gamma_\alpha \Slash P_h \gamma_\lambda) \right] \nn 
&=&-2iP_h^\mu P_h^\rho  \epsilon^{ \lambda \alpha wP_h}  \epsilon_{\lambda\alpha\rho\sigma}{\rm Tr}[\gamma^-\gamma^i \gamma^\nu \gamma^\sigma\gamma_5]  \nn
 &=& - 16 P_h^\mu P_{h\sigma} \epsilon^{-i \nu \sigma} \,.  \label{id}
\eeq
Multiplying by $A_\mu A_\nu$ and adding the $\mu \leftrightarrow \nu$ terms, we get 
\beq
&&-16\epsilon^{i j}\Bigl(P_{hj} (P_h\cdot A A^- + A^- P_h\cdot A) -P_h^-(P_h\cdot A A_j+A_jP_h\cdot A)  \Bigr) \nn
&&= -8\epsilon^{i j} z_2^2 \left(- 2\frac{k_{j}}{k^-} F^{-\mu}F^-_{\ \mu} + F^{-\mu}F_{j\mu} + F_{j\mu}F^{-\mu} -F^-_{\ j} \partial_\mu A^\mu -\partial_\mu A^\mu F^-_{\ j} \right)\,.
\eeq 
The first term gives the gluon distribution $G$, while the other terms are subleading. 
The factor in the denominator is simplified as 
\beq
\left(xp+ \left(\frac{1}{z_1}-\frac{1}{z_2}\right)P_h\right)^2 =\frac{\vec{P}_{hT}^2}{z_2} \left(\frac{1}{z_1}-\frac{1}{z_2}\right)\,.
\eeq

Next we compute the ${\mathcal O}(N_c)$ contribution.  It is easy to see that the terms proportional to $\delta^\mu_\alpha$ in the three-gluon vertex do not contribute. (Note that $\Slash k \gamma^- = \frac{\Slash P_h}{z_2} \gamma^-$.)  The terms proportional to $g_{\alpha \beta}$ can be evaluated similarly to (\ref{id}) 
\beq
&& 8P_{h\sigma} \Bigl( \epsilon^{-i \nu \sigma}(k^\mu + 2P_h^\mu (\frac{1}{z_1}-\frac{1}{z_2})) + \epsilon^{-i \mu \sigma}(k^\nu + 2P_h^\nu(\frac{1}{z_1}-\frac{1}{z_2}) ) \Bigr)A_\mu A_\nu
\\
&& =  16z_2^2\left( \frac{1}{z_1}-\frac{1}{z_2}\right) \frac{\epsilon^{ij}}{k^-} \Biggl\{ 
-k_j F^{-\mu}F^-_{\ \mu} + \frac{k^-}{2} (F_{j\mu}F^{-\mu} + F^{-\mu}F_{j\mu}) \Biggr\} 
  -8\frac{z_2^2}{z_1} \epsilon^{i j}  \left( \partial_\mu  A^\mu F^{-}_{\ j} +F^-_{\ j} \partial_\mu \cdot A^\mu \right)  \,. \nonumber 
\eeq 
 Among the terms proportional to $\delta^\mu_\beta$, only the term $2k_\alpha$ gives a nonvanishing contribution.  
Using $k_\alpha = (P_{h\alpha}-\delta^-_\alpha P_h^+)/z_2$, we get 
\beq
&& -\frac{16P_h^+}{z_2}\Bigl[ \epsilon^{j-wP_h}\bigl(P_h^- (A_j A^i +A^i A_j) -P_h^i(A_j A^-+A^- A_j)\bigr) + \epsilon^{i -wP_h}(P_h\cdot A A^- +A^- P_h\cdot A -P_h^- A^\mu A_\mu)\Bigr]\nn
&& \approx -\frac{16}{z_2}\left( \frac{1}{2}\epsilon^{jl} (F_{jl}F^{-i}+F^{-i}F_{jl})+  \epsilon^{i j}P_{hj} \frac{z_2}{k^-} F^{-\mu}F^-_{\ \mu} \right)\,.
\eeq  
The factor in the denominator is
\beq
\left(k+ \left(\frac{1}{z_1}-\frac{1}{z_2}\right)P_h\right)^2 = -\frac{\vec{P}^2_{hT}}{z_1z_2}\,.
\eeq

All in all, (\ref{last}) becomes 
\beq
&&  {\rm Im} {\rm Tr} [\Delta_F^i(z_1,z_2)(S_\alpha^L(z_1,z_2)+S_\alpha^R(z_1,z_2))] = -g^2M_N\frac{(2\pi)^4}{4z_2}  h_1(x) S_{T i}  \frac{ {\rm Im}\hat{E}_F (z_1,z_2)}{N_c}  \nn &&  \qquad \times \Biggl\{ \frac{8z_2^2}{P_T^2}\epsilon^{i j}P_{hj} \frac{G(x_g,k_T)}{N_c^2-1} \left(N_c^2+\frac{1}{ z_1\left(\frac{1}{z_2}-\frac{1}{z_1}\right)}\right) - \frac{4z_2^3}{P_T^2}\epsilon^{i j}\langle\partial\cdot A F^-_{\ j}+F^-_{\ j}\partial \cdot A\rangle \Biggr\}  \,. \label{43}
\eeq
where we omitted higher twist terms. We kept the gauge-dependent terms just to note that the prefactor $1/(N_c^2-1)$ has been canceled so that they have the same $N_c$-dependence  as the other gauge dependent terms in (\ref{fi}). Below we shall omit them because they are also subleading.

\subsection{Comparison to the fully collinear result}

Summing (\ref{cro}) and (\ref{43}), we finally obtain, relabeling $z_2\to z$,  
\beq
 E_h\frac{d\sigma^{frag}}{ d^3P_{h}} &=& 
\frac{M\alpha_s \pi^2}{N_c s}S_{T i} \epsilon^{i j} \int \frac{dz}{z^2}h_1(x) \Biggl\{ -  {\rm Im}\,  \tilde{e}(z)    \frac{d}{d(P_h^j/z)} G\left(x_g=\frac{P_{hT}^2}{xz^2s}, \frac{P_{hT}}{z}\right)   \nn 
&& + 4P_{hj} \int \frac{dz_1}{z_1^2 }\frac{z}{\frac{1}{z}-\frac{1}{z_1}} \frac{ {\rm Im}\hat{E}_F (z_1,z)}{N_c^2-1} \frac{G(x_g,P_{hT}/z)}{P_{hT}^2}  \left(N_c^2+\frac{1}{ z_1\left(\frac{1}{z}-\frac{1}{z_1}\right)}\right)  \Biggr\} \,.
\label{me}
\eeq
Let us check if (\ref{me}) is consistent with the result previously obtained in the collinear  twist-three framework relevant in the high-$P_{hT}$ region \cite{Metz:2012ct}. 
At large $P_{hT}\gg \Lambda_{QCD}$, we can use 
\beq
 G(x_g,P_{hT}/z) \approx \frac{N_c\alpha_s}{\pi^2}\frac{z^2}{x_gP_{hT}^2}  \int dx'G(x') =  \frac{N_c\alpha_s}{\pi^2}\frac{xz^4 s}{P_{hT}^4}  \int dx'G(x')  \,,
\eeq
 where $G(x)$ is the usual collinear gluon distribution.
 (\ref{me}) reduces to 
\beq
E_h\frac{d\sigma^{frag}}{ d^3P_{h}}
&=& 4M\alpha_s^2 \int \frac{dz}{z^3}  h_1(x)   \epsilon^{i j}  S_{T i}P_{hj} \frac{xz^6  }{(P^2_{hT})^3} \int dx'G(x')   \nn 
&& \qquad \times \left\{  - {\rm Im}\,  \tilde{e}(z)  + \int \frac{dz_1}{z_1^2 }\frac{1}{\frac{1}{z}-\frac{1}{z_1}} \frac{ {\rm Im}\hat{E}_F (z_1,z)}{N_c^2-1}  \left(N_c^2+\frac{1}{ z_1\left(\frac{1}{z}-\frac{1}{z_1}\right)}\right)   \right\}\,. \label{almost}
\eeq
This should be compared with Eq.~(15) of \cite{Metz:2012ct} which uses different notations for the fragmentation functions. 
\beq
M\hat{e}_{\bar{1}}  = -M_h H\,, \qquad M_N {\rm Im}\tilde{e} = 2M_h \hat{H}\,, \qquad 
M{\rm Im}\hat{E}_F(z_1,z)=2M_h z^2 \hat{H}^{\mathfrak I}_{FU}(z,z_1)\,. 
\eeq
Taking the limit $\hat{s}\gg |\hat{t}|$ in the quark-gluon channel  ($\hat{s}=xx's$, $\hat{t}=-2\frac{x}{z}p^+P_h^- = -P_{hT}^2/z^2$ are the partonic Mandelstam variables), we find 
\beq
E_h \frac{d\sigma^{frag}}{ d^3\vec{P}_h} &=& -\frac{2\alpha_s^2M_h}{s} \epsilon^{i j}S_{T i} P_{hj} \int \frac{dz}{z^3}\int \frac{dx'}{x'} \frac{1}{x\hat{s}^2} h_1(x)G(x')  \nn 
&& \quad \times\left(\frac{H(z)}{z} + \frac{2z}{N_c^2-1}\int \frac{dz_1}{z_1^2} \frac{\hat{H}^{\mathfrak I}_{FU}(z,z_1)}{\left(\frac{1}{z}-\frac{1}{z_1}\right)^2} \right) \frac{2x\hat{s}^3}{\hat{t}^3} \nn 
&=&  4\alpha_s^2M_h  \epsilon^{i j}S_{T i} P_{hj} \int \frac{dz}{z^3}\int dx' h_1(x)G(x')  \nn 
&& \quad \times\left(-2\hat{H}(z)+\frac{2z^2}{N_c^2-1} \int \frac{dz_1}{z_1^2} \frac{\hat{H}^{\mathfrak I}_{FU}(z,z_1)}{\frac{1}{z}-\frac{1}{z_1}} \left( N_c^2 + \frac{1}{z_1}\frac{1}{\frac{1}{z}-\frac{1}{z_1}} \right)\right)\frac{xz^6}{(P_{hT}^2)^3}\,,  \label{col}
\eeq
where we used (\ref{ident}). 
This agrees perfectly with (\ref{almost}). \\

\section{Including saturation effects}

\begin{figure}[t]
\begin{center}
  \includegraphics[width=11.5cm,height=5cm]{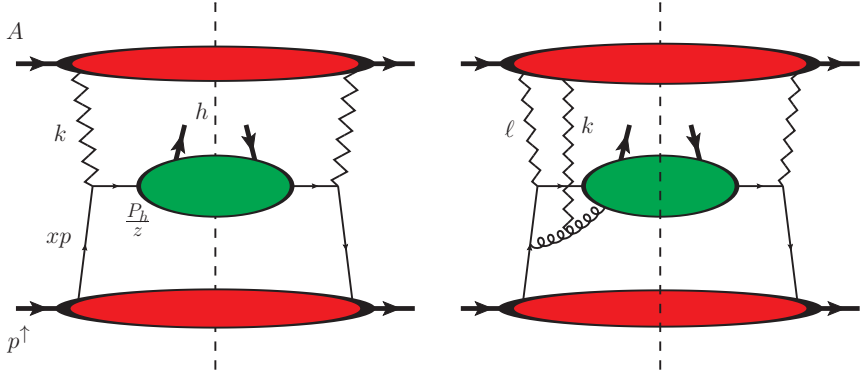}
\end{center}
 \caption{ Fragmentation contribution with saturation effects. The zigzag lines represent the multiple insertion of the $A^-$ field in the eikonal approximation. \label{fig2}  }
\end{figure}

We now include the gluon saturation effects. We closely follow the strategy used in  \cite{Hatta:2016wjz}. The diagrams to be computed are shown in Fig.~\ref{fig2}. 
The  zigzag lines  represent the Wilson line $U$ arising from the eikonal exponentiation 
\beq
ig\gamma^\mu A^a_\mu (k)t^a \to \gamma^+ \int \frac{d^2\vec{x}_T}{(2\pi)^3} e^{i\vec{x}_T\cdot \vec{k}_T} (U(\vec{x}_T)-1) \,, \qquad U(\vec{x}_T)=\exp\left(ig\int dx^+ A_a^-(x^+,\vec{x}_T)t^a \right)\,.
\eeq
In the high energy limit, the unpolarized target can be viewed as a highly Lorentz contracted shockwave. The multiple scatterings (the zigzag lines) between the polarized proton and the target can only occur either before or after the collinear gluon splitting. This is why we only need to consider the two diagrams as shown in Fig.~\ref{fig2}. 

The unintegrated gluon distribution $G(x_g,k_T)$ is converted to the correlation function of Wilson lines
\beq
\frac{x_g G(x_g,k_T)}{k_T^2} &\to& \frac{N_c}{2\pi^2\alpha_s} \int \frac{d^2x_T d^2y_T}{(2\pi)^2} e^{i\vec{k}_T\cdot (\vec{x}_T-\vec{y}_T)}  \frac{\langle q|  \frac{1}{N_c}{\rm Tr} [U^\dagger(\vec{y}_T)U (\vec{x}_T)] |q\rangle }{\langle q|q\rangle}  \nn
 &\equiv & \frac{N_c}{2\pi^2\alpha_s} F(x_g,k_T)\,, \label{rep}
\eeq
where $\langle q|q\rangle = 2q^- (2\pi)^3\delta^{(3)}(0) = 2q^- \int dx^+ d^2\vec{x}_T$. Evaluated at $x_g=\frac{P_{hT}^2}{xz^2s}$, (\ref{rep}) becomes
\beq
G\left(x_g,\frac{P_{hT}}{z}\right) \to \frac{xs N_c}{2\pi^2 \alpha_s} F\left(x_g,\frac{P_{hT}}{z}\right)\,.
\eeq
In the derivative term of (\ref{me}) which now comes from the left diagram of Fig.~\ref{fig2},   it is enough to make this replacement.  
The genuine twist-three terms are more complicated because they involve an extra collinear gluon which can be dressed by the Wilson line as shown in the right diagram of Fig.~\ref{fig2}. Still, the topology of the diagram is very similar to the one considered in \cite{Hatta:2016wjz}. We find that their color structures are exactly the same and read  
\beq
&&  \int \frac{d^2\vec{x}_T d^2\vec{y}_T d^2\vec{z}_T}{(2\pi)^6} \, (2\pi)^2\delta(\vec{k}_T+\vec{\ell}_T-\vec{P}_{hT}/z_2) e^{i\vec{k}_T\cdot \vec{z}+i\vec{\ell}_T \cdot \vec{x}_T -i\frac{\vec{P}_{hT}}{z_2} \cdot \vec{y}}
\nn 
&& \qquad \times \left\langle {\rm Tr} [U^\dagger(\vec{y})U(\vec{z})] {\rm Tr}[U^\dagger(\vec{z})U(\vec{x})]  -\frac{1}{N_c} {\rm Tr} [ U^\dagger(\vec{y})U(\vec{x})] \right\rangle
 \nn 
&&\approx  \langle q|q\rangle \delta^{(2)}(\vec{k}_T+\vec{\ell}_T-\vec{P}_{hT}/z_2) \left(  \frac{N_c^2}{ \int d^2\vec{x}}  F(x_g,\ell_T)  -\delta^{(2)}(\vec{k}_T)\right) F(x_g,P_{hT}/z_2)\,, \label{nis}
\eeq
 where we used the large-$N_c$ approximation  in the nonlinear term
\beq
\langle q| {\rm Tr} [U^\dagger(\vec{y})U(\vec{z})] {\rm Tr}[U^\dagger(\vec{z})U(\vec{x})] |q\rangle \approx \frac{\langle q| {\rm Tr} [U^\dagger(\vec{y})U(\vec{z})] |q\rangle\langle q| {\rm Tr}[U^\dagger(\vec{z})U(\vec{x})] |q\rangle}{\langle q|q\rangle}\,.  \label{nc}
\eeq


We now compute the hard part. There are two propagator denominators
\beq
\int d\ell^- \frac{1}{\left((\frac{P_h}{z_1}-\ell)^2+i\epsilon\right) \left( (xp+\ell-\frac{P_h}{z_1})^2 +i\epsilon\right)} \,.
\eeq
The two poles in $\ell^-$ are located in the opposite sides of the real axis because $\hat{E}_F(z_1,z_2)$ has a support at $z_1>z_2$ \cite{Kanazawa:2013uia}. 
We pick up the pole at $(\frac{P_h}{z_1}-\ell)^2=0$ at which
\beq
\frac{1}{\left(xp+\ell -\frac{P_h}{z_1}\right)^2}= -\frac{z_2}{z_1\left(\frac{\vec{P}_{hT}}{z_1}-\vec{\ell}_T\right)^2}\,.
\eeq
As for the numerator, we only need to calculate the component $\mu=\nu=+$. 
\beq
\epsilon^{ \lambda \alpha wP_h}  
 {\rm Tr}\left [ i\gamma_5 \sigma^{-i} \gamma^+ \gamma_5 \Slash P_h \gamma_\lambda \gamma^+ \left(\frac{\Slash P_h}{z_1}-\Slash \ell\right)\gamma^\beta \right]   \left( - 2g_{\alpha \beta}\left(\frac{1}{z_1}-\frac{1}{z_2}\right)P_h^+ + 2\delta^+_{\beta}k_\alpha \right) 
\approx  \frac{32(P_h^+)^2}{z_1}\epsilon^{ij} \left(\frac{P_{hj}}{z_1}-\ell_j\right)\,. \nonumber
\eeq
We thus arrive at the product 
\beq
&& \int d^2 \vec{k}_T d^2 \vec{\ell}_T \frac{\frac{P_{hj}}{z_1}-\ell_j  }{ \left( \frac{\vec{P}_{hT}}{z_1}-\vec{\ell}_T\right)^2} \delta^{(2)}\left(\vec{k}_T+\vec{\ell}_T-\frac{\vec{P}_{hT}}{z_2}\right) \left( \frac{N_c^2}{ \int d^2\vec{x}_T}  F(x_g,\ell_T) -\delta^{(2)}(\vec{k}_T) \right) F(x_g, P_{hT}/z_2) \nn 
&& =  \int  d^2 \vec{\ell}_T \frac{\frac{P_{hj}}{z_1}-\ell_j  }{ \left( \frac{\vec{P}_{hT}}{z_1}-\vec{\ell}_T\right)^2} \left( \frac{N_c^2}{ \int d^2\vec{x}_T}  F(x_g,\ell_T)  -\delta^{(2)}(\vec{\ell}_T-\frac{\vec{P}_{hT}}{z_2}) \right) F(x_g,P_{hT}/z_2)\,. \label{red}
\eeq
In the dilute limit, $F(x_g,\ell_T) \to \delta^{(2)}(\vec{\ell}_T)\int d^2\vec{x}_T$, and (\ref{red}) correctly reduces to  the combination in (\ref{almost})
\beq
z_1 \frac{ P_{hj}  }{ \vec{P}^2_{hT}}  \left( N_c^2 +\frac{ 1}{z_1\left(\frac{1}{z_2}-\frac{1}{z_1}\right) } \right) F(x_g,P_{hT}/z_2) \,.
\eeq
In the general case, we can perform the angular integral
\beq
\int d^2\vec{\ell}_T \frac{\frac{P_{hj}}{z_1}-\ell_j  }{ \left( \frac{\vec{P}_{hT}}{z_1}-\vec{\ell}_T\right)^2}F(\ell_T) =  2\pi z_1\frac{P_{hj}}{P_{hT}^2} \int_0^{P_{hT}/z_1}  \ell_T d\ell_T F(\ell_T)\,,
\eeq
and obtain
\beq
&& E_h\frac{d\sigma^{frag}}{ d^3\vec{P}_{h}} = 
\frac{M}{ 2 }S_{T i} \epsilon^{i j} \int  \frac{dz}{z^2} xh_1(x) \Biggl\{   -{\rm Im}\,  \tilde{e}(z)   \frac{d}{d P_h^j/z}   F\left(x_g,\frac{P_{hT}}{z} \right)  \label{fina} \\
&& \quad +4\frac{P_{hj}}{P_{hT}^2} \int_{z}^\infty \frac{dz_1}{z_1^2 }\frac{z}{\frac{1}{z}-\frac{1}{z_1}} \frac{ {\rm Im}\hat{E}_F (z_1,z)}{N_c^2-1}  \left(\frac{2\pi N_c^2}{\int d^2\vec{x}_T} \int_0^{P_{hT}/z_1} \ell_T d\ell_T F(x_g,\ell_T)  +\frac{1}{z_1\left(\frac{1}{z}-\frac{1}{z_1}  \right) } \right)F(x_g,P_{hT}/z)  \Biggr\} \,. 
\nonumber
\eeq
This is the main result of this paper.
If we assume the form 
\beq
F(x_g,\ell_T) = \frac{\int d^2\vec{x}_T}{\pi Q_s^2} e^{-\ell_T^2/Q_s^2}\,, \label{qs}
\eeq
which is a good approximation when $\ell_T^2 \leq Q_s^2$, we get 
\beq
\frac{2\pi N_c^2}{\int d^2\vec{x}_T} \int_0^{P_{hT}/z_1} \ell_T d\ell_T F(\ell_T)=
N_c^2\left(1-e^{-\frac{P^2_{hT}}{z^2_1  Q^2_s}} \right) \,. \label{su}
\eeq
Thus the effect of saturation is to reduce the ${\mathcal O}(N_c^2)$ contribution for $P_{hT}<z_1Q_s$. 

\section{Discussion}
The total spin-dependent cross section in the saturation regime is the sum of (\ref{fina}) and the soft gluon pole contribution calculated in \cite{Hatta:2016wjz}
\beq
E_h \frac{d\sigma^{SGP}}{ d^3\vec{P}_{h}} &=& -\frac{\pi Mx_F}{2(N_c^2-1)} \epsilon^{ij}S_{Ti} \int_{x_F}^1 \frac{dz}{z^3}D(z) \Biggl\{ - \frac{1}{(P_{hT}/z)^2} \frac{\partial}{\partial P_h^j/z} \left( \frac{P^2_{hT}}{z^2} F(x_g,P_{hT}/z) \right) G_F(x,x)  
\nonumber \\ 
&& \qquad+ \frac{ 2P_{hj}/z}{(P_{hT}/z)^2} F(x_g,P_{hT}/z) x \frac{d}{dx}G_F(x,x) \Biggr\} \,, \label{sat}
\eeq 
 where $G_F(x,x)$ is the Qiu-Sterman function \cite{Qiu:1991pp}. (As shown in \cite{Hatta:2016wjz}, the contribution from the soft fermionic pole vanishes in the saturation region.)
Note that in (\ref{sat}) the $P_h^j$-derivative acts on  $P_{hT}^2$ times   $F$, not $F$ itself as in (\ref{fina}).

Let us discuss the phenomenological implications of our result. Consider the dependence of the asymmetry $A_N$ on the atomic mass number $A$. In the $k_T$-factorization approach, one only has the Collins-like term proportional to ${\rm Im}\tilde{e}\sim \hat{H}$ in (\ref{fina}).  Assuming the form (\ref{qs}), one gets
\beq
\frac{\partial}{\partial P_h^j}F   \sim \frac{P_{h}^j}{Q_s^2} F\,,
\eeq
at low momentum $P_{hT}<Q_s$. Since $Q_s^2 \propto A^{1/3}$, one finds that $A_N\propto A^{-1/3}$, namely, the asymmetry is suppressed in $pA$ collisions. This is essentially the result of  \cite{Kang:2011ni}. Turning to the other  terms in (\ref{fina}) proportional to ${\rm Im}\hat{E}_F$, we see that the ${\mathcal O}(N_c^2)$ term scales as 
\beq
\frac{P_{hj}}{P_{hT}^2} \left(1-e^{-\frac{P^2_{hT}}{z^2_1  Q^2_s}} \right) \sim \frac{P_{hj}}{Q_s^2}
\eeq
 for $\Lambda_{QCD} \ll P_{hT}\ll Q_s$. Therefore, this term also leads to the behavior $A_N\sim A^{-1/3}$.  On the other hand, the ${\mathcal O}(N_c^0)$ term has a different $P_{hT}$ dependence $\sim P_{hj}/P_{hT}^2$ which implies $A_N \sim A^0$. However,  a recent study  \cite{Kanazawa:2014dca} suggests that this term is numerically small compared to the other terms in (\ref{fina}). We thus conclude that $A_N$ from the twist-three fragmentation functions (\ref{fina}) scales as $A_N \sim A^{-1/3}$ in the forward region at low momentum  $\Lambda_{QCD} \ll P_{hT}\ll Q_s$.  
This is in contrast to the observation in  \cite{Hatta:2016wjz}  that $A_N$ from the ETQS function (\ref{sat}) is independent of $A$. Indeed, the dominant term in the forward region is expected to be the derivative term $x\frac{d}{dx} G_F(x,x)$. Since its coefficient is proportional to $P_{hj}/P_{hT}^2$, we get $A_N\sim A^0$.

Experimentally, the preliminary STAR data   \cite{Heppelmann:2016siw}
 show that $A_N$ is almost independent of $A$ at least up to $x_F=0.7$. This favors the interpretation that SSA is dominated by the derivative term in (\ref{sat}). However, such an interpretation is inconsistent with the recent fit to the $p^\uparrow p \to hX$ data   
 in \cite{Kanazawa:2014dca}. There it was concluded that neither the Sivers nor Collins contribution extracted from the SIDIS data is sufficient to explain the observed  asymmetry. To resolve this problem, the authors assumed that  the genuine twist-three function ${\rm Im}\hat{E}_F \sim \hat{H}_{FU}^{\mathfrak I}$, not previously constrained by any data, is large. In particular,  the term proportional to $N_c^2$ in (\ref{fina}) was found to be the dominant   contribution.  
Yet,  our result (\ref{su}) shows that this term is most strongly affected by the saturation effect and, as we have just argued, gives rise to the scaling $A_N\sim A^{-1/3}$. We thus think more work and more data are needed to finally pin down the origin of SSA in QCD.

\acknowledgements
We thank Yuji Koike and Daniel Pitonyak for discussions.  This research was supported by the U.S. Department of Energy, 
Office of Science, Office of Nuclear Physics, under contract number 
DE-AC02-05CH11231, and by the NSFC under Grant No.~11575070. S.~Y. is supported  by the U.S. Department of Energy, Office of Science under Contract No. DE-AC52-06NA25396 and the LANL LDRD Program.

\end{document}